\documentclass[twocolumn,showpacs,preprintnumbers,amsmath,amssymb]{revtex4}
\usepackage{bbm}

\usepackage{amsfonts}
\usepackage{mathrsfs}
\usepackage{txfonts}
\usepackage{amsmath}
\usepackage{amstext}
\usepackage{graphicx}
\usepackage{dcolumn}
\usepackage{bm}

\newif\ifdvi
\ifx\pdfoutput\undefined
 \ifdvi
    \usepackage[dvips]{hyperref}
 \else
    \usepackage[dvipdfm]{hyperref}
\fi
  \AtBeginDvi{} 
\else
  \usepackage[pdftex]{hyperref}
\fi
\hypersetup{pdfstartview=FitH,%
            CJKbookmarks=true,%
       bookmarksnumbered=true,%
              colorlinks=true,%
               linkcolor=blue,
               urlcolor=blue,%
               citecolor=blue,%
              plainpages=false%
              }

\begin{document}

\title{Deterministic remote preparation of arbitrary photon polarization states}

\author{Wei Wu}
 \email{weiwu@nudt.edu.cn}
\author{Wei-Tao Liu}%
\author{Ping-Xing Chen }
\author{Cheng-Zu Li}
\affiliation{Department of Physics, National University of Defense Technology, Changsha, 410073,
People's Republic of China}

\date{\today}

\begin{abstract}
We propose a deterministic remote state preparation scheme for photon polarization qubit states, where entanglement, local operations and classical communication are used. By consuming one maximally entangled state and two classical bits, an arbitrary (either pure or mixed) qubit state can be prepared deterministically at a remote location. We experimentally demonstrate the scheme by remotely preparing 12 pure states and 6 mixed states. The fidelities between the desired and achieved states are all higher than 0.99 and have an average of 0.9947.
\end{abstract}

\maketitle

\section{introduction}
Quantum information science brings us into a whole new era, so that the information can be
manipulated and processed with quantum mechanical systems. One of the remarkable exhibitions of the
fascination of quantum information science is quantum teleportation \cite{teleportation}, which can
transmit an unknown state from one location to another without sending a physical copy of the
initial state. Remote state preparation (RSP), which is another significant application of
entanglement, has been proposed recently \cite{lorsp, patirsp, bennettrsp}. Unlike teleportation,
however, in RSP Alice (the sender) knows completely the desired state. Alice is supposed to help
Bob (the receiver) prepare the desired state at a remote location with the aid of her complete
knowledge of the desired state, prior shared entanglement and classical communications.

In recent years, RSP has attracted much attention, and various approaches towards RSP have been
studied experimentally with varying degrees of control over remotely prepared qubits. Using
liquid-state NMR, remote preparation of pseudopure states is experimentally realized firstly
\cite{NMRrsp}. Since then, the experimental remote preparation of several kinds of constrained
states have also been reported \cite{rsp2, rsp3, rsp4, rsp5}. RSP can even be realized with
classical correlations instead of quantum correlations (i.e., entanglement) \cite{ccsrsp}.
Recently, arbitrary remote control of single-qubit state have been experimentally realized
\cite{petersrsp, prsp, liursp, atomicrsp}. In Ref. \cite{petersrsp}, the trigger photon and the
remote photon are entangled in a Bell state, thus projection measurement on the trigger photon in a
basis which contains the desired state will project the remote photon into the desired state or a
state orthogonal to the desired state. Due to the impossibility of a universal NOT operation on
arbitrary qubit states \cite{NOT}, the efficiency for remote preparation of pure states are only
50\%. The efficiencies for remote preparation of mixed states depends on the desired state,
and are at least 50\%. The efficiency can be 100\% only if the desired state is constrained to
lie on a single great circle on the Poincar\'{e} sphere. In Ref. \cite{prsp}, the efficiencies for
remote preparation of arbitrary qubit states (including pure states and mixed states) also depend on the desired state, which are at least 50\%. In Ref. \cite{atomicrsp, liursp}, both
the polarization and the spatial mode of the photon are considered. Polarization beam splitter
(PBS) acts as CNOT logic gate between the polarization (control qubit) and spatial (target qubit)
for complete Bell-state measurement \cite{BSM1, BSM2, BSM3}. The efficiencies for remote
preparation of arbitrary pure states in Ref. \cite{atomicrsp, liursp} are 100\% at the cost of
precisely controlling \emph{two} interferometers \cite{note}. The efficiency for remote
preparation of mixed states in Ref. \cite{liursp} remain 50\% owing to the impossibility of a
universal NOT operation. Thus far, to our best knowledge, there is \emph{no} RSP implementation
which realize remote preparation of arbitrary single-qubit states (including \emph{pure} states
and \emph{mixed} states) deterministically.

In this paper, we report the first experimental demonstration of deterministic remote preparation
of arbitrary single-photon polarization states, where entanglement, local operation and classical
communications (LOCC) are employed. By virtue of positive operator-valued measures (POVM), we can
realize deterministic remote preparation of arbitrary pure states at a cost of one entanglement
bit (ebit) and two classical bits (cbits). By combining POVM and controlled decoherence, we can
also achieve deterministic remote preparation of arbitrary mixed states. The communication
costs are the same as that in remote preparation of pure states. Furthermore, instead of two
Mach-Zehnder interferometers in Ref. \cite{atomicrsp, liursp}, only \emph{one} interferometer is
needed in our scheme. This kind of simplification makes our scheme more feasible and executable in possible practical applications. In order to evaluate the performance of our scheme, we remotely prepare 12 pure
states and 6 mixed states. The fidelities between the desired and achieved states are all higher than 0.99 and have an average of 0.9947.

\section{Theoretical Protocol}
\subsection{Deterministic implementation of arbitrary POVM on single-photon polarization state}
POVM on single-photon polarization state plays a crucial role in our RSP protocol. So it would be the
best to start from the deterministic realization of arbitrary POVM on single-photon polarization state with linear optics elements.

POVM is the most general class of quantum measurement \cite{kraus}, which can be described by a
collection of operators $\{M_{m}\}$. The subscript $\textbf{\emph{m}}$ labels the possible
measurement outcomes. If the system state to be measured is described by a density matrix $\rho$,
then after the measurement the system state becomes
\begin{equation}\label{rhom}
\rho_{m} = \frac {M_{m} \rho M_{m}^{\dagger}}{tr(M_{m}^{\dagger} M_{m} \rho) },
\end{equation}
and the corresponding probability is given by $p_{m} = tr(M_{m}^{\dagger} M_{m} \rho)$. The
measurement operators $\{M_{m}\}$ satisfy the completeness equation $\sum_{m} M_{m}^{\dagger} M_{m}
=I$, where $I$ is unit matrix. If we define that $E_{m} \equiv M_{m}^{\dagger} M_{m}$, then $E_{m}$
will be a positive operator and $\sum_{m} E_{m} =I$. The operators $E_{m}$ are called \emph{POVM
elements} of the measurement and the complete set $\{ E_{m} \}$ is called a \emph{POVM}
\cite{quantuminformation}.

\begin{figure}[t]
\includegraphics[width=0.4\textwidth]{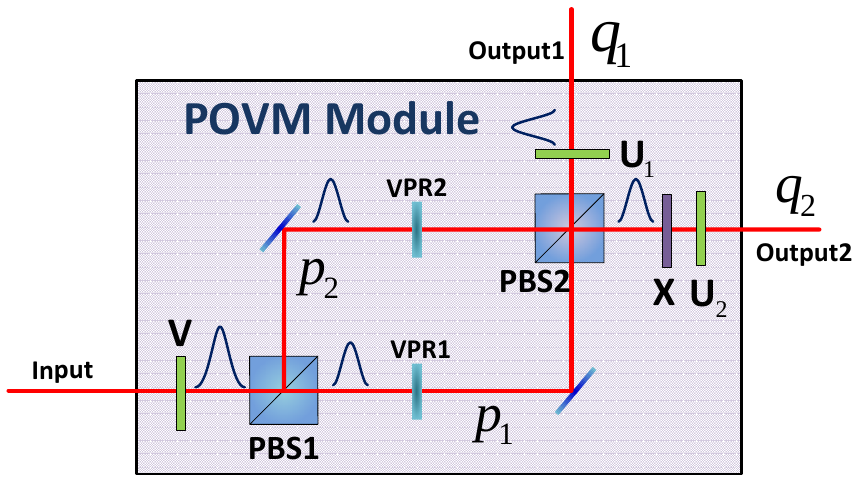}
\caption{\label{POVMmodule} Schematic diagram of the module which can implement arbitrary
two-outcome POVM on single-photon polarization state. PBS: polarizing beam splitters; VPR: variable
polarization rotator; X: NOT operator. $V$, $U_{1}$ and $U_{2}$ are variable unitary
operators.}
\end{figure}

As discussed in Ref. \cite{ET}, the module sketched in Fig. \ref{POVMmodule} can be used to
implement arbitrary two-outcome POVM on single-photon polarization state. The main part of the
module is an interferometer consisted of two polarizing beam splitters (PBS) and the relative phase between two arms is zero. Two variable polarization rotators (VPR) in the interferometer controls the polarization state in path state $|p_{1} \rangle$ or $|p_{2}\rangle$ respectively. The module also contains unitary operator $V$ at the entrance of the interferometer, unitary operator $U_{1}$ at the exit $q_{1}$ and unitary operator
$U_{2}$ plus \emph{NOT} operator $X$ at the exit $q_{2}$.
Consider the case where $V=U_{1}=U_{2}=I$, and the polarization states in path state $|p_{1}
\rangle$ and $|p_{2} \rangle$ are rotated as follows:
\begin{eqnarray}\label{VPR}
|H\rangle & \xrightarrow {VPR1} & \cos  \zeta |H\rangle + \sin
\zeta e^{i\theta}|V\rangle; \nonumber\\
|V\rangle &\xrightarrow {VPR2} & \cos  \xi |H\rangle + \sin  \xi e^{i\sigma}|V\rangle.
\end{eqnarray}
If a state in the form of $|\varphi\rangle = a | H \rangle + b| V \rangle ( |a|^2+|b|^2=1 )$ enters
the module shown in Fig. \ref{POVMmodule}, the state evolves as
\begin{eqnarray}\label{evolution}
& |\varphi\rangle \xrightarrow {} & (a \cos  \zeta |H\rangle
+ b \sin  \xi e^{i\sigma}|V\rangle) |q_{1} \rangle \nonumber\\
& & + (a \sin \zeta e^{i\theta}|H\rangle + b \cos  \xi |V \rangle) |q_{2} \rangle.
\end{eqnarray}
If one measures the output states, the output of $|q_{1} \rangle$ and $|q_{2} \rangle$ correspond
to matrices $D_{1}$ and $D_{2}$ respectively:
\begin{equation}\label{Dmatrix}
D_{1}=\left(\begin{array}{cc}\cos  \zeta & 0 \\ 0 &  \sin  \xi e^{i\sigma}
\end{array} \right),
D_{2}=\left(\begin{array}{cc}\sin \zeta e^{i\theta} & 0
\\ 0 & \cos  \xi
\end{array} \right).
\end{equation}
Note that $ D_{1}^{\dagger} D_{1} + D_{2}^{\dagger} D_{2}=I $, so when $V= U_{1} = U_{2} = I$ any
two-outcome POVM described by $D_{1}$ and $D_{2}$ can be realized with this module.

As we know, any square matrix A has its \emph{singular value decomposition}. That means there exist
unitary matrices $U$ and $V$, and a diagonal matrix $D$ with non-negative entries such that $
\emph{\textbf{A=UDV}}$. The diagonal elements of D are called the \emph{singular values} of $A$
\cite{quantuminformation}. So we represent the measurement operators $\{M_{1}, M_{2} \}$ of
arbitrary two-outcome POVM as: $\label{POVM operator} M_{1} = U_{1}D_{1}V_{1}, M_{2} =
U^{\prime}_{2}D_{2}V_{2}$. The moduli of the elements of the diagonal matrix $D_{1}$ or $D_{2}$ are confined
to lie between 0 and 1. As required by the completeness equation $E_{1} + E_{2} = I$,
\begin{eqnarray}\label{same V2}
E_{1} + E_{2} &=& V_{1}^{\dagger}D_{1}^{\dagger}D_{1}V_{1} +
V_{2}^{\dagger}D_{2}^{\dagger}D_{2}V_{2}\nonumber\\
&=& V_{1}^{\dagger}D_{1}^{\dagger}D_{1}V_{1} +
V_{2}^{\dagger}(I-D_{1}^{\dagger}D_{1})V_{2}\nonumber\\
&=& I + (V_{1}^{\dagger}D_{1}^{\dagger}D_{1}V_{1} - V_{2}^{\dagger}D_{1}^{\dagger}D_{1}V_{2})= I.
\end{eqnarray}
From Eq. (\ref{same V2}), it is easy to prove that $V_{2}= W V_{1}$ where $W$ is only a diagonal
unitary matrix. Notice that $W$ is commute with diagonal matrix $D_{2}$, so if we choose $V=V_{1}$
in the entrance, $U_{1}$ operator in the exit $q_{1}$ and $U_{2}=U^{\prime}_{2}W$ in the exit $q_{2}$, we can
implement arbitrary collection of operators $\{M_{1}, M_{2} \}$,
\begin{equation}\label{Dmatrix}
M_{1}=U_{1}\left(\begin{array}{cc}\cos\zeta & 0 \\ 0 &  \sin\xi e^{i\sigma}
\end{array} \right)V,
M_{2}=U_{2}\left(\begin{array}{cc}\sin\zeta e^{i\theta} & 0
\\ 0 & \cos  \xi
\end{array} \right)V.
\end{equation}
It means the module shown in Fig. \ref{POVMmodule} can be used to realize arbitrary two-outcome
POVM.

The realization of POVM in our module is deterministic rather than probabilistic. And any more
complicated POVM may be implemented by making a cascade of such modules. Our design is similar to
that in Ref. \cite{Ahnert2005}, however the complexity of the experimental setup is significantly
reduced, which makes it easier to realize as shown in our experiment.

\subsection{Deterministic RSP scheme for pure states}

\begin{figure}[b]
\includegraphics[width=0.48\textwidth]{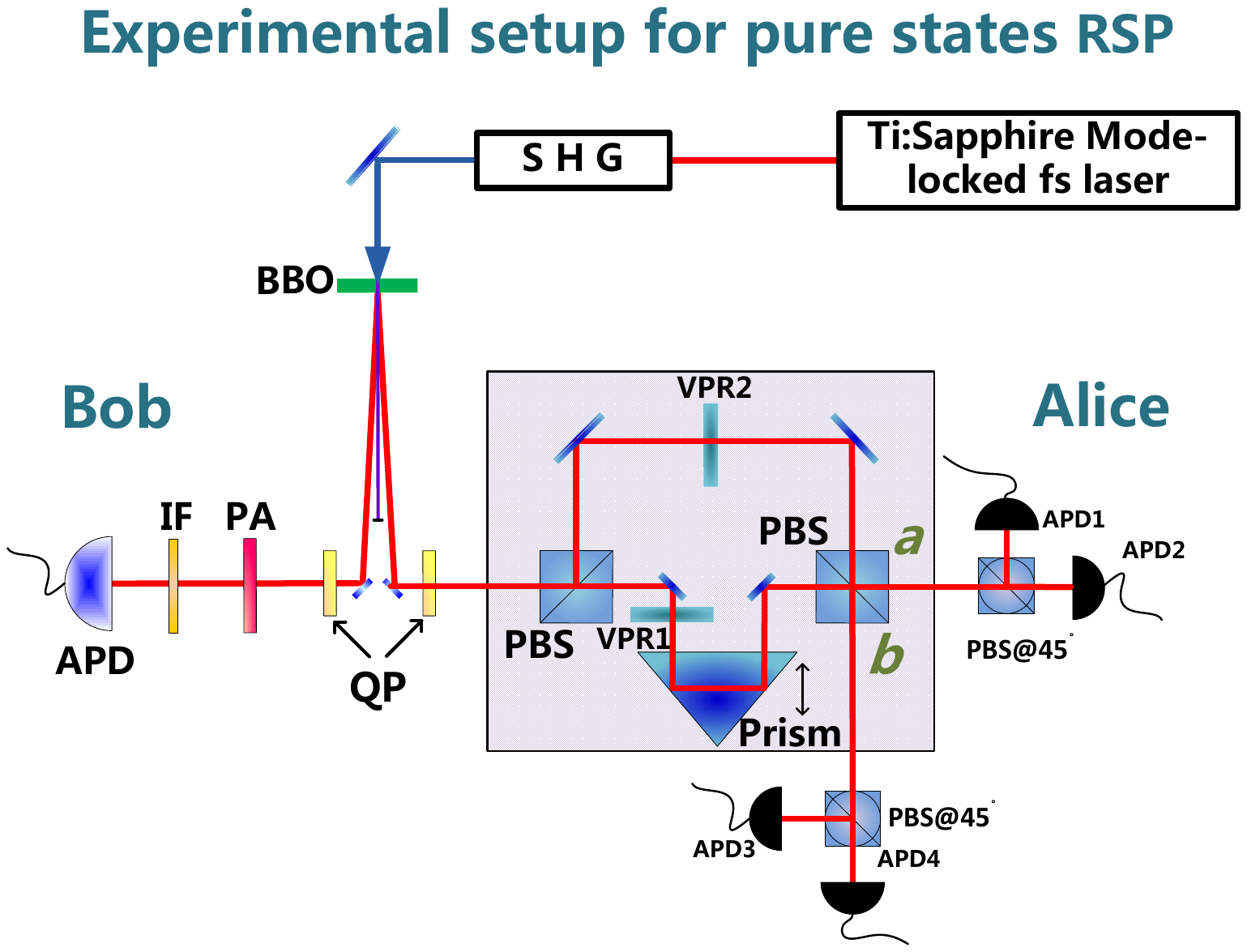}
\caption{\label{purersp} Experimental arrangement for remote preparation of pure states. SHG:
second-harmonic generator; QP: quartz crystal; PA: polarization analyzer; IF: interference filter.}
\end{figure}
In our RSP protocol, we suppose that Alice and Bob share a maximally entangled photon pair of the
form
\begin{equation}\label{bell}
|\psi_{AB}\rangle=\frac{1}{\sqrt{2}}(|H_{A}V_{B}\rangle+|V_{A}H_{B}\rangle),
\end{equation}
where the subscripts (A,B) label Alice and Bob, $|H\rangle$ and $|V\rangle$ label horizontal and
vertical polarization states of photons.

We start from remote preparation of pure states. Consider that the desired pure state is
\begin{equation}\label{stateb}
|\varphi_B\rangle=\alpha|H_B\rangle + \beta e^{i\phi}|V_B\rangle.
\end{equation}
Without loss of generality, we assume that $\alpha, \beta, \phi$ are real numbers,
$\alpha^{2}+\beta^{2}=1$ and $\phi \in [0,2 \pi)$. The experimental arrangement for remote
preparation of pure states is sketched in Fig. \ref{purersp}. VPR1 and VPR2 are arranged to rotate
the polarization component as follows:
\begin{eqnarray}\label{VPR121}
|H\rangle & \xrightarrow {VPR1} & \alpha |H\rangle + \beta e^{i\phi}|V\rangle \nonumber\\
|V\rangle & \xrightarrow {VPR2} & \alpha |H\rangle + \beta e^{i\phi}|V\rangle.
\end{eqnarray}

Then the POVM module in the shadowed box implement POVM described by:
\begin{equation}\label{ETM}
M_{1} = \left(\begin{array}{cc} \alpha & 0 \\ 0 & \beta e^{i\phi}
\end{array} \right),
M_{2} = \left(\begin{array}{cc} \beta e^{i\phi} & 0 \\ 0 & \alpha
\end{array} \right).
\end{equation}

After the POVM measurement, the initial entangled state (\ref{bell}) becomes
\begin{subequations}\label{outputpure}
 \begin{align}
 &\alpha |H_{A}V_{B}\rangle+ \beta e^{i\phi}|V_{A}H_{B}\rangle\label {outputpure1} \\
or \ \ \  &\alpha |H_{A}H_{B}\rangle + \beta e^{i\phi}|V_{A}V_{B}\rangle, \label {outputpure2}
\end{align}
\end{subequations}
depending on the measurement outcome (i.e., from which output port of the module Alice's photon flies
out).

The whole two-photon state now can be read as
\begin{eqnarray}\label {output1}
|\Psi_{AB}\rangle=&\frac{1}{2}&[|D_A^{b}\rangle(\alpha|H_B\rangle + \beta e^{i\phi}|V_B\rangle)\nonumber\\
&+&|A_A^{b}\rangle(\alpha|H_B\rangle - \beta e^{i\phi}|V_B\rangle)\nonumber\\
&+&|D_A^{a}\rangle(\alpha|V_B\rangle + \beta e^{i\phi}|H_B\rangle)\nonumber\\
&+&|A_A^{a}\rangle(\alpha|V_B\rangle - \beta e^{i\phi}|H_B\rangle)],
\end{eqnarray}
where $|D\rangle \equiv (|H\rangle + |V\rangle)/ \sqrt{2}$, $ |A\rangle \equiv (|H\rangle -
|V\rangle)/ \sqrt{2}$ and the superscripts (a,b) label the output ports \emph{a} and
\emph{b} (see Fig. \ref{purersp}).

The PBS($@45^{o}$) and the photon detectors (APD1-4) on Alice's side fulfill the polarization
projection measurement in the basis $\{|D_A\rangle,|A_A\rangle\}$ (see Fig. \ref{purersp}).
Thus when Alice's photon is projected onto $\langle D| (\langle A|)$, Bob's photon is remotely
prepared in one of the four states which is the desired state or a state up to an elementary
correction operator. According to Alice's measurement outcomes, Bob performs local unitary
operation $\hat{I}$, $\hat{\sigma}_{z}$, $\hat{\sigma}_{x}$ or $\hat{\sigma}_{y}$ to obtain the
desired state. Tuning three parameters in Eq. (\ref{VPR121}), arbitrary pure states can be remotely
prepared deterministically. The classical information cost is 2 cbits with four possible
results.

\subsection{Deterministic RSP scheme for mixed states}
\begin{figure}[t]
\includegraphics[width=0.48\textwidth]{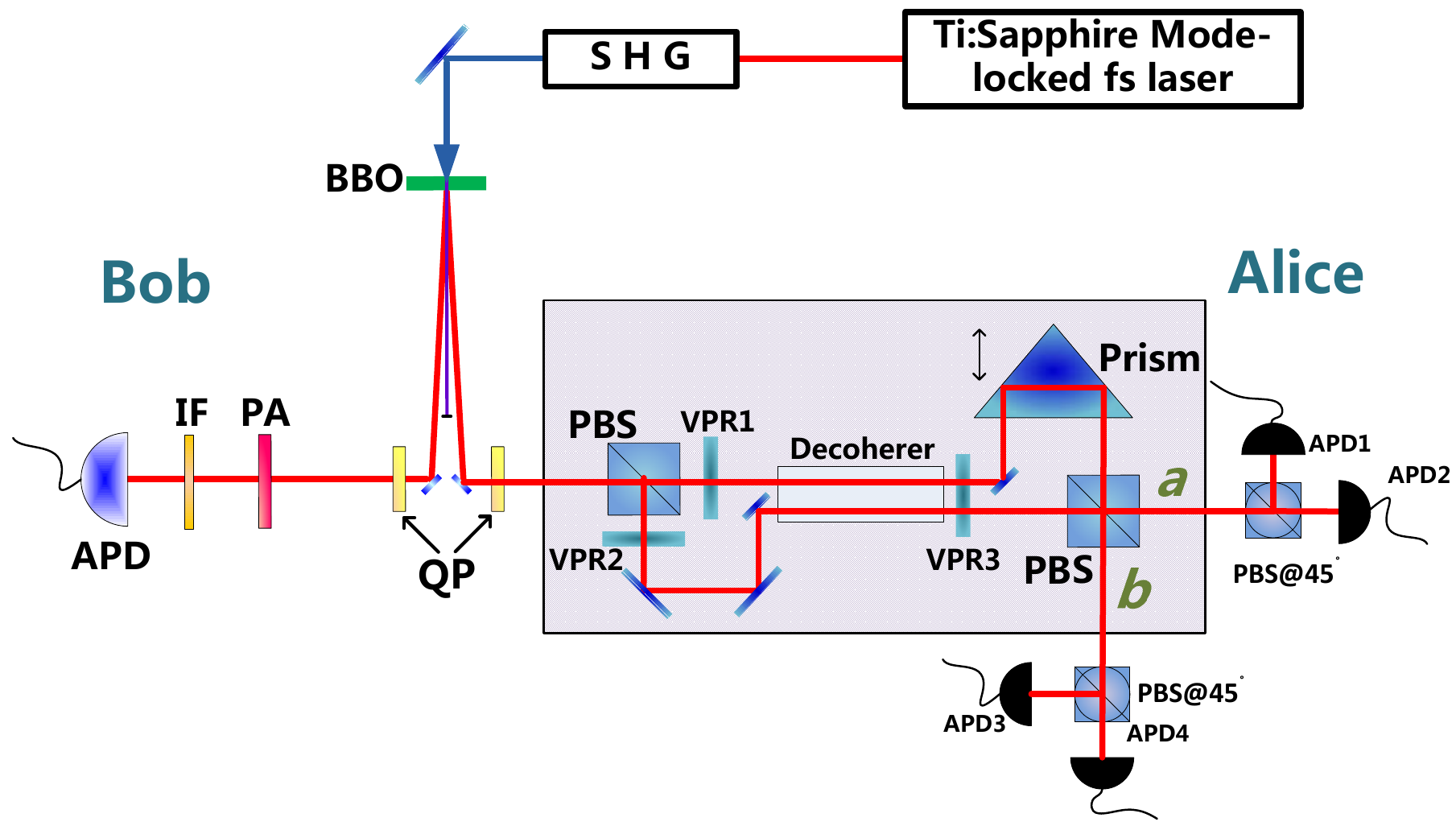}
\caption{\label{mixedrsp} Experimental arrangement for remote preparation of mixed states. SHG:
second-harmonic generator; QP: quartz crystal; PA: polarization analyzer; IF: interference filter.}
\end{figure}

Combined with POVM that allows us to remotely prepare arbitrary pure states deterministically, controlled
decoherence allows us to realize deterministic remote preparation of arbitrary mixed states. The
experimental arrangement for remote preparation of mixed states is sketched in Fig. \ref{mixedrsp},
which is the same as that in Fig. \ref{purersp} apart from the additional VPR and the decoherer.

Consider that the desired mixed state is
\begin{equation}\label{mixed}
\rho_{B}= p^{2}|\varphi_B\rangle\langle \varphi_B| + q^{2}|\varphi^{\bot}_B\rangle\langle
\varphi^{\bot}_B|\nonumber\\
\end{equation}
with
\begin{equation}\label{mixed1}
 |\varphi_B\rangle=\alpha|H_B\rangle + \beta e^{i\phi}|V_B\rangle,\ \ \
|\varphi^{\bot}_B\rangle=\beta e^{-i\phi}|H_B\rangle - \alpha|V_B\rangle,
\end{equation}
and $\langle\varphi_B|\varphi^{\bot}_B\rangle=0$. Without loss of generality, we assume that $p,
q$ are real numbers, $p^{2}+q^{2}=1 $, and $\alpha, \beta, \phi$ are the same as before. To prepare arbitrary mixed states we need to achieve complete control over all five parameters. VPR1
and VPR2 shown in Fig. \ref{mixedrsp} are arranged to rotate the polarization component as follows:
\begin{eqnarray}\label{VPR12}
|H\rangle & \xrightarrow {VPR1} & p |H\rangle + q|V\rangle \nonumber\\
|V\rangle & \xrightarrow {VPR2} & p |H\rangle + q|V\rangle.
\end{eqnarray}

A 20-mm-long quartz rod is inserted into both arms of the interferometer. With the fast axis of the
quartz rod oriented horizontally, the birefringent element introduces $\sim650 fs$ delay between
the V-polarized component and the H-polarized component, which is larger than the photon's
coherence time (given by $\lambda^{2}/(c\cdot\Delta\lambda) \sim240 fs$ in our experiment). VPR3 is arranged to rotate the polarization component in both arms as follows:
\begin{eqnarray}\label{VPR3}
|H\rangle & \xrightarrow {VPR3} & \alpha |H\rangle + \beta e^{i\phi}|V\rangle \nonumber\\
|V\rangle & \xrightarrow {VPR3} & \beta e^{-i\phi}|H\rangle - \alpha|V\rangle.
\end{eqnarray}
Then POVM measurement described by $M_{1}$ and $M_{2}$ are performed on both the V-polarized component and the H-polarized component.

In principle, the states can be distinguished by the different arrival time of
the photon with different polarization. However, the effective coincidence window used in the
experiment is $\sim 1ns$, which is much more larger than the time delay between the distinguishable
states ($\sim650 fs$). In this way, we trace over the timing information during state detection to
erase coherence between these distinguishable states, which is equivalent to irreversible
decoherence \cite{ericsson, decoherence}. Thus, we finally obtain the polarization-entangled mixed
state
\begin{subequations}\label{outputmixed}
 \begin{align}
 & p^{2}|\psi_{1}\rangle_{AB}\langle \psi_{1}|
+q^{2}|\psi_{3}\rangle_{AB}\langle \psi_{3}|\label {mixedoutput1} \\
or \ \ \  &p^{2}|\psi_{2}\rangle_{AB}\langle \psi_{2}| +q^{2}|\psi_{4}\rangle_{AB}\langle \psi_{4}|
\label {mixedoutput2}
\end{align}
\end{subequations}
depending on POVM measurement outcome, with
\begin{eqnarray}\label{mixed11}
 |\psi_{1}\rangle_{AB}&=&\alpha |H_{A}V_{B}\rangle+\beta e^{i\phi}
|V_{A}H_{B}\rangle\nonumber\\
 |\psi_{2}\rangle_{AB}&=&\alpha |H_{A}H_{B}\rangle+\beta
e^{i\phi} |V_{A}V_{B}\rangle\nonumber\\
 |\psi_{3}\rangle_{AB}&=&\beta e^{-i\phi}
|H_{A}V_{B}\rangle-\alpha
|V_{A}H_{B}\rangle\nonumber\\
 |\psi_{4}\rangle_{AB}&=&\beta e^{-i\phi}
|H_{A}H_{B}\rangle-\alpha |V_{A}V_{B}\rangle.\nonumber
\end{eqnarray}

Then the PBS($@45^{o}$) and the detectors on Alice's side perform the same projection measurement
as before, which projects Bob's photon onto one of the four mixed states:
\begin{eqnarray}\label{mixedB}
\hat{\rho}_{B}^{I}&=& p^{2}|\varphi_B\rangle\langle \varphi_B| +
q^{2}|\varphi^{\bot}_B\rangle\langle \varphi^{\bot}_B|\nonumber\\
\hat{\rho}_{B}^{X}&=& p^{2}(\hat{\sigma}_{x}|\varphi_B\rangle)(\langle \varphi_B|\hat{\sigma}_{x})
+
q^{2}(\hat{\sigma}_{x}|\varphi^{\bot}_B\rangle)(\langle \varphi^{\bot}_B|\hat{\sigma}_{x})\nonumber\\
\hat{\rho}_{B}^{Y}&=& p^{2}(\hat{\sigma}_{y}|\varphi_B\rangle)(\langle \varphi_B|\hat{\sigma}_{y})
+
q^{2}(\hat{\sigma}_{y}|\varphi^{\bot}_B\rangle)(\langle \varphi^{\bot}_B|\hat{\sigma}_{y})\nonumber\\
\hat{\rho}_{B}^{Z}&=& p^{2}(\hat{\sigma}_{z}|\varphi_B\rangle)(\langle \varphi_B|\hat{\sigma}_{z})
+ q^{2}(\hat{\sigma}_{z}|\varphi^{\bot}_B\rangle)(\langle \varphi^{\bot}_B|\hat{\sigma}_{z}),
\end{eqnarray}
So Bob obtain the desired mixed state or a mixed state up to an elementary correction operator.
According to Alice's measurement outcome, Bob performs local unitary operation $\hat{I}$,
$\hat{\sigma}_{z}$, $\hat{\sigma}_{x}$ or $\hat{\sigma}_{y}$ to achieve the desired mixed state.
Notice that an arbitrary mixed states can be remotely prepared by
tuning five parameters in Eq. (\ref{mixed}) and the classical communication required is 2 bits.

\section{experiment and results}
Our initial states (\ref{bell}) are generated with spontaneous parametric down conversion (SPDC).
As shown in Fig. \ref{purersp} and Fig. \ref{mixedrsp}, a 1-mm-thick $\beta$-barium borate (BBO)
crystal is pumped by UV laser pulses with 425 nm center wavelength and $\sim$ 530 mW average power
from a frequency-doubled mode-locked Ti:Sapphire laser with $\sim$ 200 fs pulse duration and 76 MHz
repetition rate. The photons obtained in degenerate, non-collinear type-II phase matching SPDC
process are prepared in the state of Eq. (\ref{bell}) after the quartz crystals compensate the
birefringence effects in BBO \cite{kwiat95}. We perform Clauser-Horne-Shimony-Holt (CHSH) inequality
test on the entangled state and find that $S=2.6640\pm 0.0103$($|S| \leq 2$ for any local realism theory) \cite{CHSH}.

For both pure and mixed states, PBS($@45^{o}$) at the output ports of the POVM module are used to
preform projection measurement on Alice's photon in the basis $\{|D\rangle,|A\rangle\}$. The photons are detected by single photon counting avalanche photodiode (SAPD) (Perkin-Elmer, SPCM-AQR-16) after an interference filter(10 nm FWHM). Coincidence (within a 1 ns time window) between Bob's photon and corresponding trigger photon
serves as classical communication. The coincidence circuit consists of a time-to-amplitude
converter, a single-channel analyzer (TAC\textbackslash SCA, ORTEC 567) and a universal time
interval counter (Stanford Research Systems, SR620).

In our experiment, high visibility and long stable duration of the interferometer are crucial to
the achievement of high fidelities. As shown in Fig. \ref{purersp} and Fig. \ref{mixedrsp}, the
prism is utilized to compensate the path length difference (i.e., the relative phase) between two
arms of the interferometer. The motor stage loading the prism is an ultra-precision linear motor
stage (Newport, XMS50), and the resolution is below 1 nm which is precise enough for the compensattion
of the path length difference.

The interferometer is located in a box fixed on an air cushion table to reduce the phase
fluctuation. The visibility of the interferometer can maintain above 97\% for several minutes which makes it possible to accomplish the whole tomography process and obtain high fidelities. The relative phase between two
arms should keep being zero during the remote preparation process, otherwise the fidelity would
dramatically decreased. So before the remote preparation of pure states, we insert a polarizer
($@45^{\circ}$) at the entrance of the POVM module. If the relative phase adjusted by the prism
is set to be zero, the output state of the POVM module should be $|\varphi\rangle = \alpha
|H\rangle + \beta e^{i\phi}|V\rangle$. The polarization analyzer (PA) at the output ports are used to
perform projection measurement on the output polarization state in a basis $\{|\varphi\rangle,
|\varphi^{\perp}\rangle\}$. If the visibility $
(N_{\varphi}-N_{\varphi^{\perp}})/(N_{\varphi}+N_{\varphi^{\perp}})$ is near 100 \%, we can make
sure that the relative phase is near zero and the POVM module preforms the POVM operators in
Eq.(\ref{ETM}). Because the interferometer can maintain high visibility for several minutes, now we
take off the polarizer and set the PA to measure in the basis $\{|D\rangle,
|A\rangle\}$, then the stabilization time left is enough for the qubit tomography at Bob' side. The
manipulation of the POVM module in remote preparation of mixed states is similar. In our experiment
the interferometer was not actively stabilized, however, we believe that active stabilization of
the interferometer should be employed in practical applications.

In remote preparation of mixed states, the time delay introduced in one arm should
be as exactly same as that in another arm. So that we can make sure that POVM measurement are accurately performed on both the foregoing H-polarized component and the following V-polarized component. To guarantee this, we use one quartz rod instead of two quartz rods to introduce the time delay on both arms (see Fig. \ref{mixedrsp}), which avoid the length disagreement between any two quartz rods due to the manufacturing tolerance. Then both polarization components can perfectly interfere simultaneously and measurement
operators $\{M_{1}, M_{2} \}$ can be performed precisely on both components.

\begin{figure}[t]
\includegraphics[width=0.33\textwidth]{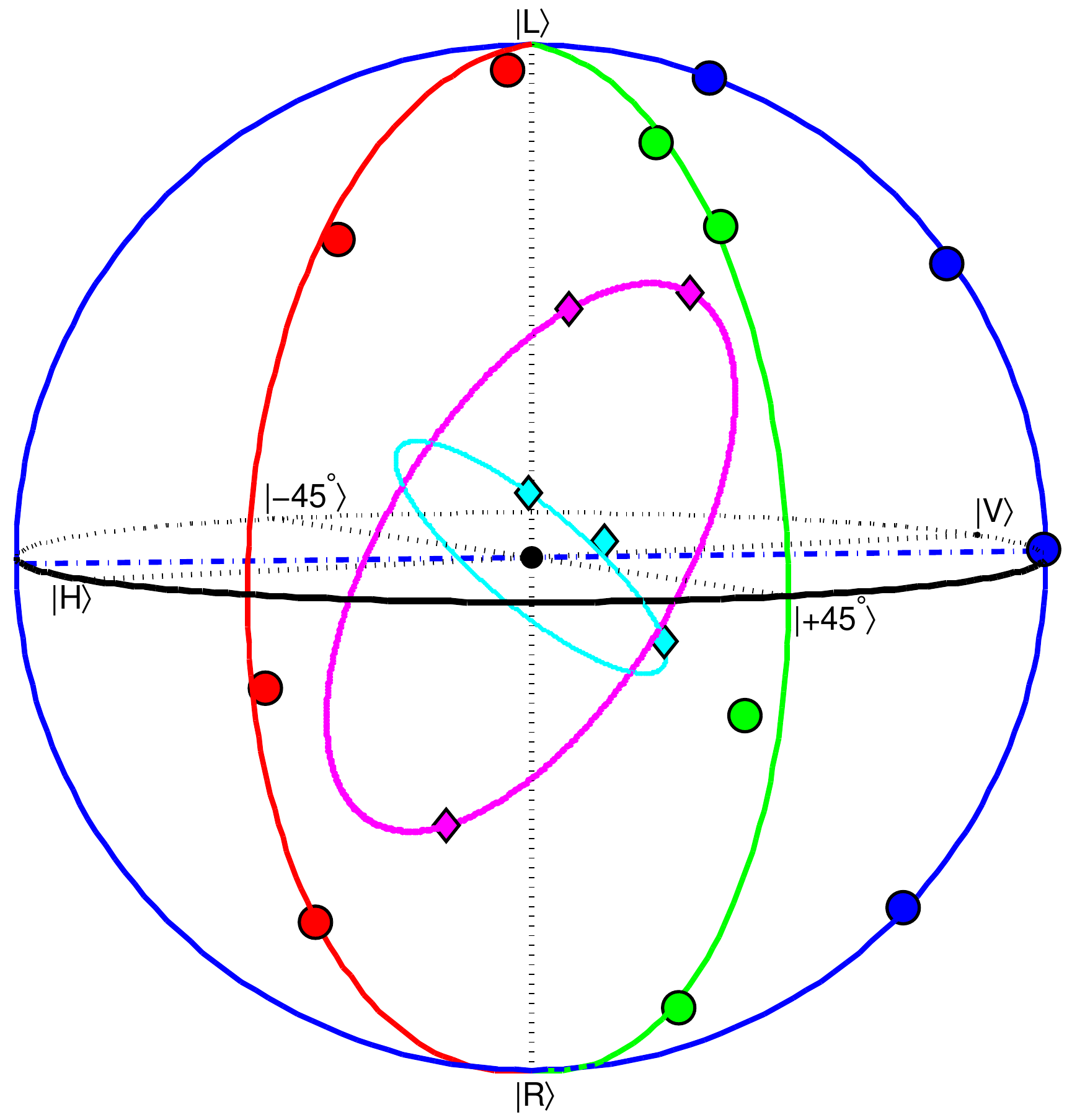}
\caption{\label{ADRSPpoincare} States remotely prepared in our experiment are shown in the
Poincar\'{e} sphere. States are supposed to lie on the (semi)circle with the same color. The pure
states are marked by circle and the mixed states are marked by diamond.}
\end{figure}

To evaluate the performance of our deterministic preparation scheme, we prepared 18 different
states on Bob's photon which include four pure states along each of three random longitude of the
Poincar\'{e} sphere and six mixed states in the Poincar\'{e} sphere (see Fig. \ref{ADRSPpoincare}).
With the tomography system on Bob's side, we perform complete polarization analysis on the prepared
polarization states. The results are converted to the closest physically valid density matrix using
a maximum likelihood technique \cite{tomography}. We use the fidelity
$F(\rho_{o},\rho_{B})\equiv|Tr(\sqrt{\sqrt{\rho_{B}}\rho_{o}\sqrt{\rho_{B}}})|^2$ to evaluate the
agreement between the prepared state ($\rho_{o}$) and the desired state ($\rho_{B}$)
\cite{fidelity}. The mean fidelity over all 18 states with all four possible results is 0.9947 in
our experiment, while F=1 means perfect match. And the fidelities of all states are above 0.99.

\section{conclusions and discussions}

In our experiment, POVM are employed to achieve deterministic remote preparation of arbitrary
photon polarization states. In fact, the kernel of our scheme is the entanglement transformation
from the initial state (\ref{bell}) to the two-photon output state (\ref{outputpure}) or
(\ref{outputmixed}). Once the desired entanglement transformation is realized deterministically, we
just need to perform appropriate projection measurement on Alice's photon and the remote
preparation is accomplished deterministically, as shown in our experiment.

Although we discuss the qubits encoded in the polarization of photons in our scheme, the methods
can be generalized to other situations. While photons are ideal carriers in transfer of qubits,
the matter carrier (e.g., ions, atoms, quantum dots, or superconducting circuits) are especially
suitable for storage and processing of qubits. The operations on Alice's photon can be utilized to
remote control other matter systems provided that the matter system is maximally entangled with
Alice's photon \cite{atomicrsp}, which is valuable for future applications such as quantum repeater
and quantum networks.

 We experimentally demonstrate the scheme by remotely preparing 12 pure states and 6 mixed states. The fidelities between the desired and achieved states are all higher than 0.99 and have an average of 0.9947.

In conclusion, we propose a deterministic remote state preparation scheme for photon polarization qubit states, where entanglement, local operations and classical communication are used. An arbitrary qubit state can be prepared deterministically at a remote location by consuming one maximally entangled state and two classical bits. The fidelities between the desired and prepared states are all higher than 0.99 and have an average of 0.9947, which indicate the high reliability of our protocol. Moreover, the experiment arrangement is more compact than before with only one interferometer used, which makes it more feasible and executable in further practical applications of quantum information science.

\acknowledgements We would like to thank Xiongfeng Ma for helpful discussion and constructive suggestion. The work is supported by National Natural Science Foundation of China (No. 10774192) and A Foundation for the Author of National Excellent Doctoral Dissertation of PR China (No. 200524).


\begin{thebibliography}{1}
\bibitem{teleportation} C. H. Bennett, \textit{et al.}, Phys. Rev. Lett. \textbf{70}, 1895 (1993).
\bibitem{lorsp} H. K. Lo, Phys. Rev. A \textbf{62}, 012313 (2000).
\bibitem{patirsp} A. K. Pati, Phys. Rev. A \textbf{63}, 014302 (2001).
\bibitem{bennettrsp} C. H. Bennett, \textit{et al.}, Phys. Rev. Lett. \textbf{87}, 077902 (2001).
\bibitem{NMRrsp} X.-H. Peng, \textit{et al.}, Phys. Lett. A \textbf{306}, 271 (2003).
\bibitem{rsp2} S. A. Babichev, \textit{et al.}, Phys. Rev. Lett. \textbf{92}, 047903 (2004).
\bibitem{rsp3} E. Jeffrey, \textit{et al.}, New J. Phys. \textbf{6}, 100 (2004).
\bibitem{rsp4} M. Ericsson, \textit{et al.}, Phys. Rev. Lett. 94, 050401 (2005).
\bibitem{rsp5} G.-Y. Xiang, \textit{et al.}, Phys. Rev. A \textbf{72}, 012315 (2005).
\bibitem{ccsrsp} W. Wu, \textit{et al.}, Opt. Comm. \textbf{281}, 1751 (2008).
\bibitem{petersrsp} N. A. Peters, \textit{et al.}, Phys. Rev. Lett. \textbf{94}, 150502 (2005).
\bibitem{prsp} W. Wu, \textit{et al.}, Int. J. Quantum Inf. to be published.
\bibitem{atomicrsp} W. Rosenfeld, \textit{et al.}, Phys. Rev. Lett. \textbf{98}, 050504 (2007).
\bibitem{liursp} W.-T. Liu, \textit{et al.}, Phys. Rev. A \textbf{76}, 022308 (2007).
\bibitem{NOT} V. Buzek, M. Hillery, and R. F. Werner, Phys. Rev. A \textbf{60}, R2626 (1999).
\bibitem{BSM1} S. P. Walborn, \textit{et al.}, Nature (London) \textbf{440}, 1022 (2006).
\bibitem{BSM2} Y. F. Huang, \textit{et al.}, Phys. Rev. Lett. \textbf{93}, 240501 (2004).
\bibitem{BSM3} C. Schuck, \textit{et al.}, Phys. Rev. Lett. \textbf{96}, 190501 (2006).
\bibitem{note} Due to lack of variable beam splitter, in Ref. \cite{liursp} the authors use a fixed beam splitter and apply different attenuators in the interferometer to realize the preparation of different states,
which limits the RSP to a certain subset of states and leads to reduction of the efficiency. An
implementation of the variable beam splitter is to use an additional Mach-Zehnder interferometer as
was done in Ref. \cite{atomicrsp}, with the cost of precisely controlling one more interferometer.
\bibitem{kraus} K. Kraus, \textsl{Lecture Notes: States, Effects and
Operations}. Springer, (1983).
\bibitem{quantuminformation} M. A. Nielsen, I. L. Chuang, \textsl{Quantum Computation and Quantum Information}.
Cambridge University Press, (2000).
\bibitem{ET} W. Wu, \textit{et al.}, Opt. Comm. \textbf{282}, 2093 (2009).
\bibitem{Ahnert2005} S. E. Ahnert, M. C. Payne, Phys. Rev. A \textbf{71}, 012330 (2005).
\bibitem{ericsson} M. Ericsson, \textit{et al.}, Phys. Rev. Lett. \textbf{94}, 050401 (2005).
\bibitem{decoherence} P. G. Kwiat, and B.-G. Englert, \textsl{Science and Ultimate Reality}. Cambridge University Press, (2004).
\bibitem{kwiat95} P. G. Kwiat, \textit{et al.}, Phys. Rev. Lett. \textbf{75}, 4337 (1995).
\bibitem{CHSH} J. F. Clauser, \textit{et al.}, Phys. Rev. Lett. \textbf{23}, 880 (1969).
\bibitem{tomography} D. F. V. James, \textit{et al.}, Phys. Rev. A \textbf{64}, 052312 (2001).
\bibitem{fidelity} R. Jozsa, J. Mod. Opt. \textbf{41}, 2315 (1994).
\end{thebibliography}
\end{document}